\documentstyle[12pt]{article}

\def\ga{\mathrel{\mathpalette\fun >}}
\def\fun#1#2{\lower3.6pt\vbox{\baselineskip0pt\lineskip.9pt
\ialign{$\mathsurround=0pt#1\hfil##\hfil$\crcr#2\crcr\sim\crcr}}}
\title{Leptons and Photons}
\author{L. Okun\\
ITEP, Moscow, 117259, Russia\\
Preprint ITEP No. 68-95}

\begin{document}
\maketitle

\begin{abstract}

It is assumed that three lepton families $(\nu_e, e)$, $(\nu_{\mu},
\mu)$, $(\nu_{\tau}, \tau)$ carry charges, which are sources of
electronic, muonic and tauonic massless vector particles,
respectively. Various manifestations of these hypothetical photons
are discussed.
\end{abstract}

We know six leptons: three electrically charged ones, $e, \mu, \tau$,
and three electrically neutral ones, $\nu_e, \nu_{\mu}, \nu_{\tau}$.
We know only one photon that is coupled to the electric charge of
particles. One of the most fundamental properties of the electric
charge is its conservation. There exist three different leptonic
quantum numbers: electronic for $e, \nu_e$, muonic for $\mu,
\nu_{\mu}$, and tauonic for $\tau, \nu_{\tau}$, which at the present
level of our ignorance are conserved. If they are strictly conserved
they may serve as corresponding charges emitting and absorbing
$\gamma_e, \gamma_{\mu}, \gamma_{\tau}$ -- electronic, muonic and
tauonic photons, respectively. These hypothetic charges and
hypothetic photons and their possible manifestations are the subject
of this letter.

If neutrino oscillations are discovered
experimentally, that will prove that the number of leptonic charges
is reduced.  Observation of transitions $\nu_e \leftrightarrow
\nu_{\mu} \leftrightarrow \nu_{\tau}$ would leave open only one
window -- that for the existence of total leptonic charge. In a
similar manner observation of neutrinoless  double beta decay $(2n
\to 2e + 2p)$ would mean that electronic charge is definitely not
conserved and hence there is no electronic photons. ( A non-conserved
charge cannot be a source of massless vector particles and of a
Coulomb-like potential, behaving as $1/r$).  The non-conservation
of electronic or muonic charge would be also proved by the relevant
proton decays, promised to us by  the grand unified theories.
(Note that in some of them B-L -- the difference between baryonic
 and leptonic charges is conserved.) But
today there is still plenty of room for speculations about all three
types of leptonic photons.

It should be stressed that the existence of leptonic photons does
not seem to solve any of the existing problems. In that sense they
are not needed. But who needed muon when it was discovered?

The leptonic photons  are not only not needed;
they are undesirable, as the additional Abelian gauge symmetries
connected with them would produce triangle anomalies, which without
special care  would destroy the
renormalizability of the Standard Model.

After presenting all necessary warnings let us proceed.
Analogously to the electric charge $e$ and to the fine structure
$\alpha = e^2/4\pi$, let us introduce $g_e, g_{\mu}, g_{\tau}$ and
$\alpha_e = g_e^2/4\pi$, $\alpha_{\mu} = g^2_{\mu}/4\pi$,
$\alpha_{\tau} = g^2_{\tau}/4\pi$. One of our goals would be to
establish the upper limits on  $\alpha_e, \alpha_{\mu},
\alpha_{\tau}$.

For $\alpha_e$ a straightforward limit is given by the tests of
independence of gravitational acceleration on the material of the
probe body. These tests were started many years ago by
E\"{o}tv\"{o}s \cite{1} at the accuracy $10^{-8}$. More recent tests
\cite{2, 3} have reached $10^{-12}$. The electronic "Coulomb force"
should give repulsion, proportional to the product of the numbers of
electrons in the earth (or sun) and the test body. Therefore the
variation of the acceleration with the kind of the test body atoms
would be \cite{4, 5}
$$
1 + \frac{\alpha_e}{\alpha_g} (\frac{\bar{Z}}{\bar{A}})(\frac{Z}{A})
$$
where $\alpha_g = G m_p^2 =  6\cdot 10^{-39}$ ($G$ --
the Newton gravitational constant, $m_p$ -- mass of proton),
$\bar{Z}$ and $\bar{A}$ are the  numbers of electrons and nucleons in
an average atom of  the earth $(\bar{Z}/\bar{A} = 0.5$) or sun
$(\bar{Z}/\bar{A}  = 1)$; while $Z$ and $A$ refer to an atom of a
test body. If we compare two balls, one of copper, another of  lead,
the difference in acceleration would be $$
\frac{\alpha_e}{\alpha_g}(\frac{\bar{Z}}{\bar{A}})(\frac{Z_{Cu}}{A_{Cu}}
- \frac{Z_{Pb}}{A_{Pb}}) < 10^{-12}
$$
Taking into account that $\frac{Z_{Cu}}{A_{Cu}} = 0.46\;, \;\;
\frac{Z_{Pb}}{A_{Pb}} = 0.40$ we get
$$
\alpha_e \leq 10^{-49}\;.
$$

This upper limit is obtained under assumption that the electronic
charge of the sun (or earth) is not neutralized by antileptons. Such
neutralization is impossible if the only light (and electrically
neutral) antileptons are antineutrinos \cite{6, 7}.
The number of $\bar{\nu}$'s
cannot be equal to that of $e$'s inside a celestial  body. Because
of Pauli principle the momenta of $\bar{\nu}$'s are of the order of
electron momenta. Thus their kinetic energy is much larger  than that
of electrons, while the potential energy is smaller by a factor
$\alpha_e/\alpha \ll 1$. Thus, there can be no equilibrium between
equal amounts of electrons and antineutrinos.

In principle, one can think \cite{6}, that the electronic charge of
the earth is screened by some neutral leptonic bosons, carrying the
electronic (leptonic) charge, such particles were discussed in ref.
\cite{8}. In ref. \cite{9, 10, 7} they were assumed to be
electronic antisneutrinos -- $\tilde{\bar
\nu}_{e}$ -- supersymmetric scalar partners of $\bar{\nu}_e$ with
$m_{\tilde{\bar{\nu}}_e} \leq 1$ eV. But for the
E\"{o}tv\"{o}s type upper limit to  become irrelevant, the screening
must be perfect.  For instance, to allow $\alpha_e \sim 10^{-20}$ the
screening must have accuracy $10^{-29}$ which does not seem
plausible.
 Especially difficult to get a substantial neutralization
of electronic charge in bodies with dimensions smaller than, say, 1
cm.  Having some of their $\tilde{\bar{\nu}}'$ outside, these
bodies may explode \cite{6, 7}. The explosion
would be caused by the leptonic repulsion of their unscreened
electrons.  Let us discuss the mechanism of explosion in some
detail for the case of no screening. Consider a sphere of radius
$r$ with density $n$ of atoms per cm$^3$ and number of electrons per
atom $Z$.  Omitting numerical coefficients we get for the force of
the volume repulsion
$$
F_R \sim \alpha_e\frac{(Znr^3)^2}{r^2} \;\; .
$$
The force of attraction of the two halves of the sphere is
proportional to the separation surface and can be estimated as
$$
F_A
= nr^2a \frac{\alpha^2}{a^2}\;\;,
$$
where $a \sim 10^{-8}$ cm is the
dimension of an atom, while $\alpha^2 /a^2$ is a crude estimate of
the Van-der-Waals force. The sphere would become unstable at $r >
r_c$.  The critical radius $r_c$ is determined by the equation
$$
F_R
= F_A \;\; ,
$$
wherefrom we obtain
$$
r_c =
(\frac{\alpha^2}{Z^2n\alpha_e a})^{1/2} \sim a \frac{\alpha}{Z
\alpha_e^{1/2}} \;\; .
$$
For $\alpha_e \sim 10^{-20}$ and $Z \sim
10$ we get $r_c \sim 1$mm.  Further theoretical work is needed to
prove or disprove  the possibility of screening of the
electronic charge of celestial bodies.

Let us now consider $\alpha_{\mu}$ and $\alpha_{\tau}$.
The upper limits for them can be derived from experiments with
$\mu$'s, $\tau$'s, and from the neutrino experiments.

An upper limit on $\alpha_{\mu}$ from the muon magnetic moment is
determined by the experimental accuracy of
$$
a_{\mu}\equiv \frac{1}{2}(g_{\mu}-2) \;\; .
$$
According to \cite{11}
$$
\frac{\Delta a_{\mu}^{exp}}{a_{\mu}}\simeq 10^{-5} \;\; .
$$

While the virtual $\gamma$ gives correction to $a_{\mu}$ close to
$\alpha/2\pi$, the virtual $\gamma_{\mu}$ gives $\alpha_{\mu}/2\pi$.
Hence $\alpha_{\mu}$ must be at least five orders of magnitude
smaller than $\alpha$. This means that about $10^{-5}$ of all photons
emitted by muons could be $\gamma_{\mu}$'s. Muonic photons would be
emitted more effectively by $\nu_{\mu}$'s, for instance in pion decays
the ratio
$$
\frac{\Gamma(\pi\to\mu\nu_{\mu}\gamma_{\mu})}{\Gamma
(\pi\to\mu\nu_{\mu})} \sim
\frac{2\alpha_{\mu}}{\pi}\frac{d\omega}{\omega} \ln
\frac{2\Delta}{m_{\nu_{\mu}}} \;\; ,
$$
where $\omega$ is the energy of the photon in the pion rest frame,
$\Delta = m_{\pi}-m_{\mu}$, and we assume $\omega \ll \Delta \ll
m_{\mu}$. Thus
$\nu_{\mu}$-beams would be accompanied by $\gamma_{\mu}$-beams.
Muonic photons would be penetrating and would produce pairs of muons
when colliding with nuclei. The cross-section of this process
on a pointlike nucleus with charge $Z$ at asymptotically high energy
$\omega$ would be \cite{12}:
$$
\sigma= \frac{\alpha^2 Z^2\alpha_{\mu}}{m_{\mu}^2}
\cdot \frac{28}{9}[ \ln \frac{2\omega}{m_{\mu}}-\frac{109}{42}]\;\;.
$$
The search for
such pairs may further decrease the upper limit on $\alpha_{\mu}$.

Similar limit on $\alpha_{\tau}$ from $(g_{\tau}-2)$ is non-existent.
Therefore a search for $\tau$-photons in $\tau$- and
$\nu_{\tau}$-experiments is even more interesting. If CHORUS and
NOMAD do not find $\nu_{\tau}$-oscillations, they may find narrow
$\mu^+\mu^-$ pairs from $\gamma_{\mu}$'s or $\tau^+\tau^-$ pairs from
$\gamma_{\tau}$'s.

Cosmological limit on the mass density of relic
$\nu_{\mu}$'s and $\nu_{\tau}$'s gives lower limits on
$\alpha_{\mu}$ and $\alpha_{\tau}$ \cite{4},
assuming that $m_{\nu_{\mu}},~m_{\nu_{\tau}}\ga~ 0.1~ KeV$ and that
these neutrinos were not burned out by some exotic processes:
$$
\alpha_{\mu} > 10^{-12}(\frac{m_{\nu_{\mu}}}{1 KeV}) \;\; , \;\;
\alpha_{\tau} > 10^{-12}(\frac{m_{\nu_{\tau}}}{1 KeV}) \;\; .
$$

These lower limits follow from the requirement that the annihilation
of the pairs $\nu_{\mu}\bar{\nu}_{\mu}$ and
$\nu_{\tau}\bar{\nu}_{\tau}$ into leptonic photons was fast enough to
guarantee an acceptable mass density of these particles.

Of special interest are the neutrino electric charges induced by the
mixing of ordinary and leptonic photons \cite{4}. The mixing appears
through the vacuum polarization loops of $e, \mu, \tau$ because these
leptons carry both leptonic and ordinary charges. The simplest
Lagrangian  of four photons without mixing may be written (omitting
the four-vector indues) in the form:
$$
{\cal{L}}=\sum_i(\frac{1}{4}F_i^2 +g_i A_i j_i) \;\; ,
$$
where $i=0$ refers to ordinary photon $\gamma$ (so that $g_0\equiv
e$), while $i =1,2,3$ refer to $e, \mu, \tau$, correspondingly.

In the one-loop approximation three terms have to be added to
${\cal{L}}$: $FF_e$, $FF_{\mu}$, $FF_{\tau}$ with coefficients
proportional to $\sqrt{\alpha\alpha_e}$, $\sqrt{\alpha\alpha_{\mu}}$
and $\sqrt{\alpha\alpha_{\tau}}$, correspondingly. In the two-loop
approximation three more terms should be added $F_e F_{\mu}$, $F_e
F_{\tau}$, $F_{\mu}F_{\tau}$ with coefficients
$\alpha\sqrt{\alpha_e\alpha_{\mu}}$,
$\alpha\sqrt{\alpha_e\alpha_{\tau}}$ and
$\alpha\sqrt{\alpha_{\mu}\alpha_{\tau}}$, correspondingly. Moreover
the coefficients of the diagonal terms $\frac{1}{4}F^2$,
$\frac{1}{4}F_e^2$, $\frac{1}{4}F_{\mu}^2$, $\frac{1}{4}F_{\tau}^4$
should be changed.

Were $\alpha_e$ not negligibly small, the induced electric charges,
of $\nu_e$ and of $e$ could be observable:
$$
Q_{\nu_e} -Q_n = Q_e -Q_p \neq 0 \;\; .
$$
As a result the electrical neutrality of atoms would be lost . (It is
tested at the level of $10^{-20}$ -- $10^{-23}$; for a collection of
experimental data see e.g. ref. \cite{13}).

The mixings $\gamma\leftrightarrow \gamma_{\mu}$ and
$\gamma\leftrightarrow \gamma_{\tau}$ will not damage the neutrality
of atoms, but will produce nonvanishing electric charges of
$\nu_{\mu}$ and $\nu_{\tau}$.

For light $\nu_{\mu}$ and $\nu_{\tau}$ ($m\ll$ 1 KeV) the upper
limits on their electric charges are given by the neutrino luminosity
of the sun \cite{14}
$$
Q_{\nu_{\mu}} ~~~~ \mbox{and} ~~~~ Q_{\nu_{\tau}}\leq 10^{-13} \;\; .
$$
Independently of the value of the mass of $\nu_{\mu}$, an upper limit
on its electric charge can be derived from the data on
$\nu_{\mu}e$-scattering \cite{16}:
$$
Q_{\nu_{\mu}}\leq 10^{-8}~~.
$$

The mixing coefficients, derived in one- or two-loop approximation
are logarithmically divergent. This divergences are cancelled by
corresponding counterterms. Thus the resulting mixing coefficients
and the neutrino electric charges might be quite different from their
naive estimates. After diagonalization of ten $F_i F_k$ terms we are
left with four diagonal "photons", but each of them will be coupled to
all four currents.

Further complications arise if we take into account the box-type muon
or tauon loops \cite{4} which contribute to the magnetic moments of
charged particles. A lot of exotic phenomena (similar to neutrino
oscillations) appear if leptonic photons are not strictly massless,
but have tiny masses \cite{16}.

Finally, let us return to the problem of anomalies.

Looking for an anomaly-free scheme with leptonic photons,
for which E\"{o}tvos type experiments are irrelevant, it is
natural to consider a
scheme with one leptonic photon which has no coupling to $(e, \nu_e)$
and has non-vanishing couplings of equal strength and opposite sign
to $(\mu, \nu_{\mu})$ and $(\tau, \nu_{\tau})$. In this case the
triangle with one leptonic photon and two electroweak gauge fields
vanishes due to cancellation between $(\mu, \nu_{\mu})$ and $(\tau,
\nu_{\tau})$, while the triangle with two leptonic photons and one
electroweak
gauge field vanishes due to cancellation between neutrino and its
charged partner. In the framework of this scheme the leptonic photons
emitted by muons in the high-energy neutrino experiments could
in principle produce
not only pairs $\mu^+\mu^-$, but also pairs $\tau^+\tau^-$.
However the cross-section of the latter process would be three orders
of magnitude smaller.

Another type of anomaly-free scheme has been suggested to me by
M.B.Vo\-lo\-shin. In his scheme the
baryo-leptonic photon is coupled to the difference of baryonic and
leptonic charges: B-L. Its coupling to all three generations has to
be universal because of the Kobayashi-Maskawa mixing between quarks.

As was pointed out to me by A.D.Dolgov, recent observational
data on light element abundances allow existence of one extra
massless particle. (See e.g. B.Fields, K.Kainulainen, K.Olive, hep-ph
9512321.)

I am grateful to A.D.Dolgov for useful discussions of the manuscript
of this letter. This research was made possible in part  by  grant
93-02-14431 of the Russian Foundation for Basic Research.  The main
content of this letter was presented at the Symposium in honour of
Klaus Winter on the occasion of his 65th birthday, on December 6,
1995 at CERN.

\end{document}